\documentclass[journal,twoside]{IEEEtran}
\usepackage{subfigure}
\usepackage{color}
\usepackage{cite}

\usepackage{cite}
\usepackage{citesort}

\usepackage{cite,graphicx,amsmath,amssymb,cite,algorithm}

\begin{document}
\newtheorem{theorem}{Theorem}
\newtheorem{corollary}{Corollary}
\newtheorem{remark}{Remark}

\title{Artificial-Noise-Aided Secure Transmission with Directional Modulation based on Random Frequency Diverse Arrays}
\author{{Jinsong Hu, Shihao Yan,~\IEEEmembership{Member,~IEEE,} Feng Shu,~\IEEEmembership{Member,~IEEE,} Jiangzhou Wang,~\IEEEmembership{Fellow,~IEEE,}\\ Jun Li,~\IEEEmembership{Senior Member,~IEEE,} and Yijin Zhang~\IEEEmembership{Member,~IEEE}}

\thanks{This work was supported in part by the National Natural Science Foundation of China (Nos. 61271230, 61472190, 61501238 and 61301107), the Open Research Fund of National Key Laboratory of Electromagnetic Environment, China Research Institute of Radiowave Propagation (No. 201500013), the open research fund of National Mobile Communications Research Laboratory, Southeast University, China (No. 2013D02), the Jiangsu Provincial Science Foundation Project (BK20150786), the Specially Appointed Professor Program in Jiangsu Province, 2015, the Fundamental Research Funds for the Central Universities (No. 30916011205), and the Australian Research Council's Discovery Projects (DP150103905).
}

\thanks{J. Hu, F. Shu, J. Li, and Y. Zhang are with the School of Electronic and Optical Engineering, Nanjing University of Science and Technology, Nanjing, China. F. Shu is also with National Key Laboratory of Electromagnetic Environment, China Research Institute of Radiowave Propagation, China, and with National Mobile Communications Research Laboratory, Southeast University, Nanjing, China. (emails: \{jinsong\_hu, shufeng, jun.li\}@njust.edu.cn; yijin.zhang@gmail.com).

S. Yan is with the Research School of Engineering, Australia National University, Canberra, ACT, Australia. (email: shihao.yan@anu.edu.au).

J. Wang is with the School of Engineering and Digital Arts, University of Kent, Canterbury CT2 7NT, U.K. (e-mail: j.z.wang@kent.ac.uk).}

}

\maketitle
\begin{abstract}
In this paper, we propose a novel directional modulation (DM) scheme based on random frequency diverse arrays with artificial noise (RFDA-DM-AN) to enhance physical layer security of wireless communications. Specifically, we first design the RFDA-DM-AN scheme by randomly allocating frequencies to transmit antennas, thereby achieving two-dimensionally (i.e., angle and range) secure transmissions, and outperforming the state-of-the-art one-dimensional (i.e., angle) phase array (PA) based DM scheme. Then we develop the closed-form expression of a lower bound on the ergodic secrecy capacity (ESC) of our RFDA-DM-AN scheme. Based on the theoretical lower bound derived, we further optimize the transmission power allocation between the useful signal and artificial noise (AN) in order to enhance the ESC. Simulation results show that 1) our RFDA-DM-AN scheme achieves a higher secrecy capacity than that of the PA based DM scheme, 2) the lower bound derived is shown to approach the ESC as the number of transmit antennas $N$ increases and precisely matches the ESC when $N$ is sufficiently large, and 3) the proposed optimum power allocation achieves the highest ESC compared with other power allocations in the RFDA-DM-AN.


\end{abstract}

\section{Introduction}

\IEEEPARstart{A}s a promising physical layer security technique, directional modulation (DM) has attracted extensive studies due to its unique characteristic. This characteristic is that DM projects modulated signals into a predetermined spatial direction while simultaneously distorting the constellation of these signals in all other directions. This can significantly decrease the probability of these signals being eavesdropped on. As such, the DM technique is an ideal candidate to achieve physical layer security \cite{yangnan2016,Trappe2015,hebiao2016,yan2014transmit,long2,yan2016location,chenxiaoming,zhaonan1}. In general, there are two main types of methods to implement the DM technique in wireless communications. The first one is to adopt DM on the radio frequency (RF) frontend (e.g., \cite{Babakhani1,Babakhani2,Daly1,Daly2}). For example, the authors of \cite{Babakhani1,Babakhani2} obtained the
phase and amplitude of DM signal at the predefined direction through varying the effective length and scattering property of a reflector. A similar approach was proposed in \cite{Daly1,Daly2}, where the phase of each antenna element was shifted accordingly in order to construct the DM signal.
However, the flexibility of implementing DM on the RF frontend is limited,
which leads to high complexity in the design of constellation diagram for DM.
Against this background, the second method was developed in the literature (e.g., \cite{Ding2}), which implemented the DM technique on the baseband instead of on the RF frontend. Specifically, the authors of \cite{Ding2} proposed an approach to apply the DM technique on the baseband based on an orthogonal vector. In addition, the authors of \cite{Jinsong,Shufeng} provided a robust baseband DM algorithm by considering estimation errors on the direction angles. Implementing DM on baseband enables dynamic DM transmissions to send the different patterns of constellation point at different time slots, which results in that tracking and decoding the useful signals is harder for an eavesdropper, and thus can further improve physical layer security.

In the literature, the DM technique can be achieved by phase array (PA)\cite{Daly1,Daly2,Ding2}. Considering security, previous studies on the DM technique only investigated the system where a legitimate user locates at the desired direction and an eavesdropper locates in another direction (that is different from the desired direction). However, in the context of physical layer security it is common to assume that the location information of the eavesdropper is not available at the transmitter, since the eavesdropper may be passive and never transmit signals, and thus it is hard, if not impossible, to obtain such location information. In practice, we may have the scenario in which the eavesdropper exactly locates in the desired direction as the legitimate user. In this scenario, the aforementioned DM based on PA can no longer guarantee the secure transmission for the legitimate user. This is due to the fact that the DM based on PA can only distort signals at the directions that are different from the desired one.

A linear frequency diverse array (LFDA) in \cite{Antonik:PHDthesis,Sammartino,wangwenqin2015} offers new possibilities for DM to guarantee a secure transmission in the aforementioned scenario where the legitimate user and eavesdropper locate in the same direction (but different ranges). This is due to the fact that LFDA can produce a beam-pattern with controllable direction and range, by linearly shifting the carrier frequencies across different transmit antennas. However, as discussed in \cite{Sammartino,wangwenqin2015}, the direction and range achieved by LFDA are coupled. This means that there may exist multiple direction-range pairs at which the eavesdropper can receive identical signals as the legitimate user, which compromises the secure transmission. Recently, the authors of \cite{Liu1,Liu2} developed a new type of frequency diverse array, namely the random frequency diverse array (RFDA), of which each transmit antenna is randomly (instead of linearly) allocated a narrow band frequency. As shown in \cite{Liu1,Liu2}, RFDA owns one property that it can decouple the correlation between the direction and range (this correlation exists in LFDA and cannot be decoupled). This property enables RFDA to be a good candidate for DM to achieve a robust secure transmission (i.e., physical layer security).
In \cite{hebiao2016}, the authors discussed two main metrics, i.e., ergodic secrecy capacity (ESC) and secrecy outage probability, which are often adopted to measure the performance of secure transmissions over fading channels. ESC applies for delay tolerant systems which allows for the adoption of an ergodic version of fading channels. On the other hand, secrecy outage probability, which measures systems with probabilistic formulations, is more appropriate for scenario under stringent delay constraints. In our work, the instantaneous value of the secrecy capacity at Eve is not available due to the strategies of randomly allocating frequencies to the transmit antennas in the proposed scheme. Averaging over all the realizations of the frequencies allocation, we can capture the ergodic features of the secrecy capacity. The concept of ESC bears the similar significance as the one adopted in this paper.

In this work, for the first time, we utilize the DM with artificial noise based on RFDA (referred to as the RFDA-DM-AN scheme) to enhance physical layer security of wireless communications. In this scheme, in addition to maximizing the signal-to-noise ratio (SNR) of useful signals at the desired direction, the transmitter also sends artificial noise (AN) in all other directions to bring interferences to the eavesdropper. In order to fully examine the secrecy performance of the RFDA-DM-AN scheme, we first derive a lower bound on its ESC. Based on this lower bound, we can determine the optimal transmit power allocation between the useful signal and AN more efficiently relative to using the ESC. As shown in our examination, this lower bound precisely matches the ESC when the number of transmit antennas is sufficiently large, which confirms the validity and effectiveness of using this lower bound to perform transmit power allocation. In addition, we investigate two strategies of randomly allocating frequencies to the transmit antennas in the RFDA (i.e., frequency allocations based on the continuous and discrete uniform distributions). Our investigation demonstrates that the continuous uniform frequency allocation outperforms the discrete one in terms of average ESC.

The remainder of this paper is organized as follows. In Section II, we detail our system model for the RFDA-DM-AN scheme. Then the secrecy performance of the RFDA-DM-AN scheme is analyzed in Section III, based on which the transmit power and frequency allocations are examined. The secrecy performance of the proposed scheme is numerically evaluated in Section IV, and Section V draws concluding remarks.

\emph{Notations:} Scalar variables are denoted by italic symbols. Vectors and matrices are denoted by lower-case and upper-case boldface symbols, respectively. Given a complex number, $|\cdot|$ and $(\cdot)^*$ denote the modulus and conjugation, respectively. Given a complex vector or matrix, $(\cdot)^T$, $(\cdot)^H$, $\mathrm{tr}(\cdot)$, and $\|\cdot\|$ denote the transpose, conjugate transpose, trace, and norm, respectively. The $N \times N$ identity matrix is referred to as $\mathbf{I}_{N}$ and $\mathbb{E}[\cdot]$ denotes expectation operation.


\section{System Model}

\subsection{Random Frequency Diverse Array}

\begin{figure}
  \centering
  \includegraphics[scale=0.75]{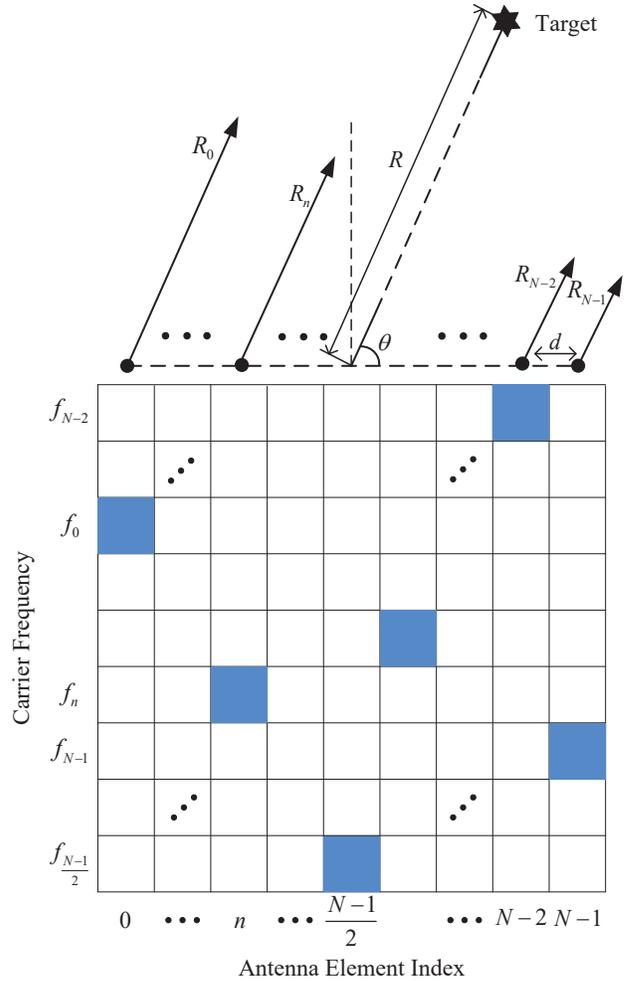}\\
  \caption{The structure of the random frequency diverse array (RFDA).}\label{The structure of the random FDA}
\end{figure}

As shown in Fig.~\ref{The structure of the random FDA}, the RFDA (i.e., random frequency diverse array) is different from the PA (i.e., phased array) due to the use of frequency increment across the antenna elements at the transmitter. The frequency allocated to the $n$-th element is given by
\begin{equation}
\label{carrier frequency of RFDA}
f_n=f_c+k_n\Delta{f},~n=0,~1,~\ldots,~N-1,
\end{equation}
where $f_c$ is the central carrier frequency and $\Delta{f}$ is the frequency increment. In the RFDA, the value of $k_n$ is normally random. One method (e.g. continuous uniform distribution) to determine the value of $k_n$, which determines one specific random mapping rule to assign the carrier frequencies of the different elements, is illustrated in Fig.~\ref{The structure of the random FDA}.
In this work, we consider a uniform linear array (ULA) at the transmitter and set the phase reference at the array geometric center. The target range for the $n$-th element is denoted as $R_n$. In practice, the target is assumed far from the antenna array, and thus $R_n$ can be approximated as
\begin{equation}
\label{range}
R_n={R-b_nd\cos{\theta}},~n=0,~1,~\ldots,~N-1,
\end{equation}
where
\begin{align}
b_n=n-\frac{N-1}{2},
\end{align}
$\theta$ and $R$ are the angle and range from target to the transmitter, and $d$ denotes the element spacing of the ULA at the transmitter. We note that in the LFDA the value $k_n$ is equal to $b_n$, which is a linear function of $n$ \cite{Antonik:PHDthesis,Sammartino}.

The phase of the transmit signal at the reference element of the ULA is given by
\begin{align}
\psi_0(\theta,R)=2\pi{f_c}\frac{R}{c}.
\end{align}
Likewise, the phase of transmit signal at the $n$-th element can be expressed as
\begin{align}
&\psi_n(\theta,R)=2\pi{f_n}\frac{R_n}{c}  \\ \nonumber
&=2\pi\left(f_c\frac{R}{c}-b_n\frac{f_c d\cos\theta}{c}+k_n\Delta{f}\frac{R}{c}-b_nk_n\Delta{f}\frac{d\cos\theta}{c}\right).
\end{align}
Then, the phase shift of the $n$-th element relative to the reference element is given by
\begin{align}\label{phase shift of RFDA}
&\Psi_n(\theta,R) \notag \\
&=\psi_n(\theta,R)-\psi_0(\theta,R) \notag \\
&=2\pi\left(-b_n\frac{f_c d\cos\theta}{c}+k_n\frac{\Delta{f}R}{c}-b_nk_n\frac{\Delta{f}d\cos\theta}{c}\right).
\end{align}
We note that the second term in~\eqref{phase shift of RFDA} is of importance, because it shows that the radiation pattern of the array depends on both the range and the frequency increment. Normally, the relationship between frequency increment and carrier frequency can guarantee $N\Delta{f}\ll f_c$, and element spacing $d$ is close to the wave length $\lambda$ (e.g. $d=\lambda/2$). As such, the third term in \eqref{phase shift of RFDA} is negligible \cite{Sammartino}. Therefore, the phase shift defined in \eqref{phase shift of RFDA} can be approximated by
\begin{align}
\Psi_n(\theta,R)\approx \frac{2\pi}{c}(-b_nf_c d\cos\theta+k_n\Delta{f}R).
\end{align}
Then, the normalized steering vector of RFDA to a specific location $(\theta, R)$ is given by
\begin{align}\label{steering_vector}
{\mathbf{h}(\theta,R)}&\!=\!\frac{1}{\sqrt{N}}[e^{j\Psi_0(\theta,R)}, e^{j\Psi_1(\theta,R)},\!\dots\!, e^{j\Psi_{N\!-\!1}(\theta,R)}]^T.
\end{align}

\subsection{Directional Modulation with Artificial Noise}

\begin{figure}
  \centering
  \includegraphics[scale=0.65]{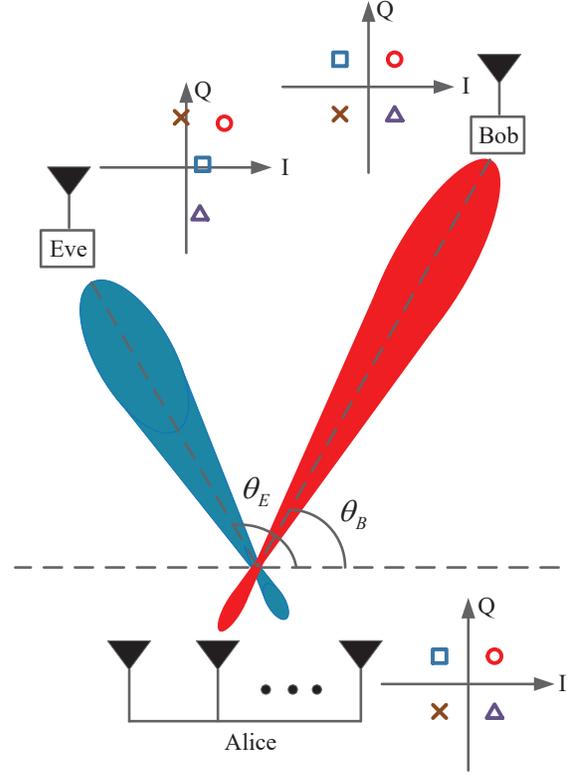}\\
  \caption{Illustration of constellation diagram in DM system for the QPSK modulation.}\label{Illustration of DM}
\end{figure}

Since DM (i.e., directional modulation) is a transmitter-side technology, this work considers a multiple-input single-output (MISO) wiretap channel as shown in Fig.~\ref{Illustration of DM}. In this wiretap channel, the transmitter (Alice) is equipped with $N$ antennas, the legitimate user (Bob) is equipped with a single antenna, and the eavesdropper (Eve) is equipped with a single antenna. We assume that the location of Bob, denoted by $(\theta_B,R_B)$, is available at Alice, while the location of Eve, denoted by $(\theta_E,R_E)$, is unavailable at Alice (which potentially exists in anywhere). In addition, we consider that free space channel model was widely adopted in the literature for the DM technique (e.g., \cite{Ding2,Jinsong,Shufeng}). Without loss of generality, we normalize the channel gain to be 1.

Beamforming with AN (i.e., artificial noise) has been widely used in the context of physical layer security due to its robustness and desirable secrecy performance \cite{Goel,yangnan2015}. Therefore, for the first time, we adopt the AN-aided secure transmission in the DM technique based on RFDA. Considering beamforming with AN, the transmitted signal can be expressed as
\begin{align}
\label{transmit signal}
{\mathbf{s}} =\sqrt{\alpha P_s}{\mathbf{v}}x+\sqrt{(1-\alpha)P_s}\mathbf{w},
\end{align}
where $x$ is a symbol chosen from the complex signal constellation with average power constraint (i.e., $\mathbb{E}[|x|^2]=1$). Also in \eqref{transmit signal}, $P_s$ is the transmit power of Alice and $\alpha$ is the parameter that determines the power allocation between the useful signal and AN. In addition, $\mathbf{v}$ is the beamforming vector for the useful signal. Since Alice does not know Eve's location, in order to maximize the SNR at Bob, $\mathbf{v}$ is given by
\begin{align}
\mathbf{v}=\mathbf{h}(\theta_B,R_B),
\end{align}
where $\mathbf{h}(\theta_B,R_B)$ is the steering vector of the RFDA at Alice to Bob, which can be obtained by replacing $(\theta, R)$ with $(\theta_B,R_B)$ in \eqref{steering_vector}.
Furthermore, the artificial noise vector $\mathbf{w}$ in \eqref{transmit signal} should lie in the null space of $\mathbf{h}(\theta_B,R_B)$ (i.e., $\mathbf{h}^H(\theta_B,R_B)\mathbf{w} = 0$) in order to avoid interference to Bob. As such, $\mathbf{w}$ can be expressed as \cite{Jinsong}
\begin{align}
\mathbf{w}=\frac{(\mathbf{I}_N-{\mathbf{h}(\theta_B,R_B)}{\mathbf{h}^H(\theta_B,R_B)})\mathbf{z}}{\Vert(\mathbf{I}_N-{\mathbf{h}(\theta_B,R_B)}{\mathbf{h}^H(\theta_B,R_B)})\mathbf{z}\Vert}.
\end{align}
where $\mathbf{z}$ consists of $N$ independent and identically distributed (i.i.d.) circularly-symmetric complex Gaussian random variables with zero-mean and unit-variance, i.e., $\mathbf{z} \sim  \mathcal{CN}(0,\mathbf{I}_N)$.

Following \eqref{transmit signal}, the received signal at Bob is given by
\begin{align}
\label{received signal of Bob}
{y(\theta_B,R_B)}&={\mathbf{h}^H(\theta_B,R_B)}{\mathbf{s}}+n_B \notag \\
&=\sqrt{\alpha P_s}{\mathbf{h}^H(\theta_B,R_B)}{\mathbf{v}}x+ n_B \notag \\
&=\sqrt{\alpha P_s}x+n_B,
\end{align}
where $n_B$ is the additive white Gaussian noise (AWGN), distributed as $n_B\sim \mathcal{CN}(0,\sigma_B^2)$. From \eqref{received signal of Bob}, we can see that Bob can restore the original signal $x$ from Alice easily without knowing the random mapping rule. To be fair, we assume that Eve cannot obtain this random mapping rule as well.
Following \eqref{received signal of Bob}, the SNR at Bob is given by
\begin{align}
\label{gamma B}
\gamma_B&=\frac{\alpha P_s}{\sigma_B^2}=\alpha\mu_B,
\end{align}
where $\mu_B=P_s/\sigma_B^2$.

Likewise, the received signal at Eve can be expressed as
\begin{align}
\label{received signal of Eve}
{y(\theta_E,R_E)}&={\mathbf{h}^H(\theta_E,R_E)}{\mathbf{s}}+n_E \notag \\
&=\sqrt{\alpha P_s}{\mathbf{h}^H(\theta_E,R_E)}{\mathbf{h}(\theta_B,R_B)}x \notag \\
&~~+\sqrt{(1-\alpha) P_s}{\mathbf{h}^H(\theta_E,R_E)}\mathbf{w}+n_E,
\end{align}
where $n_E$ is the AWGN with the distribution $n_E \sim \mathcal{CN}(0,\sigma_E^2)$ and ${\mathbf{h}(\theta_E,R_E)}$ is the steering vector of the RFDA at Alice to Eve, which can be obtained by replacing $(\theta, R)$ with $(\theta_E,R_E)$ in \eqref{steering_vector}.

As per \eqref{received signal of Eve}, we can see that the item $\sqrt{P_s}{\mathbf{h}^H(\theta_E,R_E)}{\mathbf{h}(\theta_B,R_B)}$ distorts the amplitude and phase of the signals at Eve. In addition, the item ${\mathbf{h}^H(\theta_E,R_E)}\mathbf{w}$ is nonzero since $\mathbf{h}^H(\theta_E,R_E)$ is not orthogonal with $\mathbf{w}$. This further distorts the constellation of $x$ at Eve.

Following \eqref{received signal of Eve}, the signal-to-interference-plus-noise ratio (SINR) at Eve is given by
\begin{align}
\label{gamma E}
&\gamma_E=\frac{\alpha P_s|{\mathbf{h}^H(\theta_E,R_E)}{\mathbf{h}(\theta_B,R_B)} |^2}{(1-\alpha) P_s| {\mathbf{h}^H(\theta_E,R_E)}\mathbf{w} |^2+\sigma_E^2} \notag \\
&=\frac{\alpha \mu_B|{\mathbf{h}^H(\theta_E,R_E)}{\mathbf{h}(\theta_B,R_B)} |^2}{(1-\alpha) \mu_B| {\mathbf{h}^H(\theta_E,R_E)}\mathbf{w} |^2+\beta}
\end{align}
where
\begin{align}
\beta\triangleq\frac{\sigma_E^2}{\sigma_B^2}.
\end{align}

\section{Secrecy Performance of the RFDA-DM-AN Scheme}

In this section, we analyze the secrecy performance of the RFDA-DM-AN scheme. Specifically, we first determine its ESC and then derive a lower bound on this ESC. Based on this lower bound, we determine the optimal power allocation between the useful signal and AN. Then, two strategies of randomly allocating frequencies to the transmit antennas are examined.

\subsection{Ergodic Secrecy Capacity}

In the context of physical layer security, the secrecy capacity is defined as $\{0, C_B-C_E\}^{+}$, where $C_B$ is the capacity at Bob, which is given by
\begin{align}\label{capacity of Bob}
C_B=\log_2(1+\gamma_B),
\end{align}
and $C_E$ is the capacity at Eve, which is given by
\begin{align}\label{capacity of Eve}
C_E=\log_2(1+\gamma_E).
\end{align}
We note that in the considered system model without path loss $C_B \geq C_E$ can be guaranteed, since $\gamma_B$ is the maximized by Alice.

The ESC is commonly used for the fading channel with statistical channel state information at the transmitter. In general, the ESC is defined as the instantaneous secrecy capacity averaged over $\gamma_B$ and/or $\gamma_E$. Following \eqref{gamma B}, we know that $\gamma_B$ does not depend on the frequency allocation at the RFDA (i.e., the values of $k_n$). However, following \eqref{gamma E} we note that $\gamma_E$ is a function of $k_n$ since both $\mathbf{h}(\theta_B,R_B)$ and $\mathbf{h}(\theta_E,R_E)$ are functions of $k_n$. As the distribution of $k_n$ is available at Alice (the transmitter), we adopt the ESC, which is obtained by averaging the secrecy capacity over $\gamma_E$, as the main performance metric to evaluate the secrecy performance of different schemes. We would like to mention that the randomness in $\gamma_E$ is caused by the random frequency allocation instead of the fading in our work.
Accordingly, this ESC is given by
\begin{align}
\label{ergodic capacity of RFDA}
C = \mathbb{E}\left[C_B - C_E\right] = C_B - \mathbb{E}\left[C_E\right].
\end{align}
We note that this ESC $C$ is for a specific location of Eve. However, as assumed in this work, Alice does not know Eve's location. As such, we define $\overline{C}$ as the average value of $C$ over all possible locations of Eve, which is determined by the region where Eve potentially exists. For example, the location of Eve can be assumed at an annular region centered on the location of Bob, which is similar to the annulus threat model mentioned in \cite{Shihao}. The average value of $C$ can be calculated through
\begin{align}\label{average capacity}
\overline{C}=\int_{R_E\in \mathcal{R}}\int_{\theta_E\in \Theta} C f(\theta_E,R_E) d\theta_E d R_E,
\end{align}
where $f(\theta_E,R_E)$ is the joint probability density function (pdf) of $\theta_E$ and $\mathcal{R}_E$ in the sets $\Theta$ and $\mathcal{R}$, respectively. Then, the optimal value of the power allocation parameter $\alpha$ that maximizes $\overline{C}$ can be obtained through
\begin{align}\label{optimal power allocation factor of average capacity}
\alpha^{\ast}=\underset{0\leq\alpha\leq1}{\mathrm{arg}\max}~~\overline{C}.
\end{align}

In order to efficiently determine $\alpha^{\ast}$, we have to derive a closed-form expression for $\overline{C}$. However, due to the high complexity of $\gamma_E$ as shown in \eqref{gamma E}, the closed-form expression for $C$
is mathematically intractable (not to mention the closed-form expression for $\overline{C}$). In order to facilitate the power allocation, we next derive a lower bound on the ESC $C$ in the following subsection.

\subsection{A Lower Bound on the Ergodic Secrecy Capacity}

A lower bound on the ESC $C$ is derived in the following theorem in order to facilitate the transmit power allocation between the useful signal and AN at Alice.

\begin{theorem}\label{theorem1}
The lower bound on the ESC of the RFDA-DM-AN scheme is
\begin{align}\label{lower bound of ergodic capacity}
C_{\mathrm{LB}} =\log_2\left(\frac{\!-\!\alpha^2\mu_B^2 +\alpha\mu_B(\beta F+\mu_B\!-\!1)+\beta F+\mu_B}{\alpha\mu_B(F-\frac{1}{\eta}-1)+\beta F+\mu_B}\right),
\end{align}
where
\begin{align}
&F\triangleq\frac{N^2}{\eta(N^2-N(1-\Phi^2(j2\pi p))+S^2_N(q)\Phi^2(j2\pi p))}, \\
&q\triangleq\frac{{f_c d(\cos\theta_E-\cos\theta_B)}}{c},  \\
&p\triangleq\frac{\Delta{f}(R_E-R_B)}{c}, \\
&S_N(x)\triangleq\frac{\sin(N\pi x)}{\sin(\pi x)}, \\
&\eta\triangleq1/{\mathrm{tr}\left\{\left[\mathbf{I}_N-{\mathbf{h}(\theta_B,R_B)}{\mathbf{h}^H(\theta_B,R_B)}\right]^2\right\}},
\end{align}
and $\Phi(\cdot)$ is the moment generating function (MGF) of $k_n$.
\end{theorem}

\begin{IEEEproof}
The cross correlation coefficient between $\mathbf{h}(\theta_E,R_E)$ and $\mathbf{h}(\theta_B,R_B)$ is
\begin{align}\label{cross correlation coefficient of these two steering vector RFDA}
&\mathbf{h}^H(\theta_E,R_E)\mathbf{h}(\theta_B,R_B) \notag \\
&=\frac{1}{N}\sum^{N-1}_{n=0}e^{j\frac{2\pi}{c}\{b_nf_c d(\cos\theta_E-\cos\theta_B)-k_n\Delta{f}(R_E-R_B)\}} \notag \\
&=\frac{1}{N}\sum^{N-1}_{n=0}e^{j2\pi (n-(N-1)/2)q}e^{-j2\pi k_n p}.
\end{align}

Since in the above expression of $\mathbf{h}^H(\theta_E,R_E)\mathbf{h}(\theta_B,R_B)$, only the parameter $q$, $p$, and $k_n$ are of interest since they are functions of the location information and the random frequency allocation. Then, to proceed we define
\begin{align}
\rho(q, p, k_n)\triangleq \mathbf{h}^H(\theta_E,R_E)\mathbf{h}(\theta_B,R_B).
\end{align}

The mean of $|\rho(q,p,k_n)|^2$ over $k_n$ is derived as
\begin{align}\label{mean of rho_square}
&\mathbb{E}_{k_n}[|\rho(q,p,k_n)|^2] \notag =\mathbb{E}_{k_n}[\rho^*(q,p,k_n)\rho(q,p,k_n)] \notag \\
&=\frac{1}{N^2}\mathbb{E}_{k_n,k_{n'}}\left\{\sum_{n=0}^{N-1}\sum_{n'=0}^{N-1}e^{-j2\pi [b_n q-k_n p]}e^{j2\pi [b_{n'} q-k_{n'} p]}\right\}\notag \\
&=\frac{1}{N^2}\mathbb{E}_{k_n}\left\{\sum_{n=0}^{N-1}e^{-j2\pi [b_n q-k_n p]}e^{j2\pi [b_n q-k_n p]}\right\}+\frac{1}{N^2} \notag \\
&\cdot\mathbb{E}_{k_n,k_{n'}}\left\{\sum_{n=0,n\neq n'}^{N-1}\sum_{n'=0}^{N-1}e^{-j2\pi [b_n q-k_n p]}e^{j2\pi [b_{n'} q-k_{n'} p]}\right\} \notag \\
&=\frac{N}{N^2}+\frac{1}{N^2}\Bigg\{\int_{k_n\in \mathcal{K}} g(k_n)e^{-j2\pi k_n p} dk_n\int_{k_{n'}\in \mathcal{K}} g(k_{n'})  \notag \\
&\cdot e^{-j2\pi k_{n'} p} dk_{n'}\Bigg\}\Bigg\{\sum_{n=0,n\neq n'}^{N-1}\sum_{n'=0}^{N-1}e^{-j2\pi b_n q}e^{j2\pi b_{n'} q}\Bigg\} \notag \\
&=\frac{1}{N}+\frac{1}{N^2}\Phi^2(j2\pi p)\left(\frac{\sin^2(N\pi q)}{\sin^2(\pi q)}-N\right)  \notag \\
&=\frac{1}{N^2}[N(1-\Phi^2(j2\pi p))+S^2_N(q)\Phi^2(j2\pi p)],
\end{align}
where $g(k_n)$ is the pdf of $k_n$ in the set $\mathcal{K}$.

Next, we can derive the lower bound of the ESC $C$ by using the Jensen's inequality, i.e., $\log_2 \mathbb{E}[x]\geq \mathbb{E}[\log_2 x]$. Then, following \eqref{gamma E} we have
\begin{align}\label{inequality}
C&=C_B-\mathbb{E}[C_E] \notag \\
&\geq \log_2\left(1+\alpha\mu_B\right) \notag \\
&-\log_2\left(1+\frac{\alpha \mu_B \mathbb{E}\left[|{\mathbf{h}^H(\theta_E,R_E)}{\mathbf{h}(\theta_B,R_B)} |^2\right]}{(1-\alpha) \mu_B \mathbb{E}\left[| {\mathbf{h}^H(\theta_E,R_E)}\mathbf{w} |^2\right]+\beta}\right) \notag \\
&\overset{a}{=}\log_2\left(1+\alpha\mu_B\right) \notag \\
&-\log_2\left(1+\frac{\alpha\mu_B\mathbb{E}_{k_n}[|\rho(q,p,k_n)|^2]}{(1-\alpha)\mu_B
\eta(1-\mathbb{E}_{k_n}[|\rho(q,p,k_n)|^2])+\beta}\right),
\end{align}
where $\overset{a}{=}$ is achieved by
\begin{align}
&\mathbb{E}\left[|\mathbf{h}^H(\theta_E,R_E)\mathbf{w}|^2\right]\notag\\&=\mathbb{E}\left[\mathrm{tr}\{\mathbf{h}^H(\theta_E,R_E)\mathbf{w}\mathbf{w}^H\mathbf{h}(\theta_E,R_E)\}\right] \notag \\
&=\mathbb{E}_{k_n}\Bigg[\mathrm{tr}\Bigg\{\frac{\mathbf{h}^H(\theta_E,R_E)\mathbf{P}(\theta_B,R_B)\mathbb{E}[\mathbf{z}\mathbf{z}^H]}{\mathbf{P}(\theta_B,R_B)\mathbb{E}[\mathbf{z}\mathbf{z}^H]} \notag \\
&~~~~~~~~~~~~~~~\frac{{\mathbf{P}^H(\theta_B,R_B)\mathbf{h}(\theta_E,R_E)}}{\mathbf{P}^H(\theta_B,R_B)} \Bigg\} \Bigg] \notag \\
&\overset{b}{=}\mathbb{E}_{k_n}\Bigg[\mathrm{tr}\Bigg\{\frac{\mathbf{h}^H(\theta_E,R_E)\mathbf{P}(\theta_B,R_B)\mathbf{I}_N}{\mathbf{P}(\theta_B,R_B)\mathbf{I}_N}\notag \\
&~~~~~~~~~~~~~~~\frac{{\mathbf{P}^H(\theta_B,R_B)\mathbf{h}(\theta_E,R_E)}}{\mathbf{P}^H(\theta_B,R_B)} \Bigg\} \Bigg] \notag \\
&=\frac{1-\mathbb{E}_{k_n}\left[|{\mathbf{h}^H(\theta_E,R_E)}{\mathbf{h}(\theta_B,R_B)}|^2\right]}{\mathrm{tr}\{[\mathbf{I}_N-{\mathbf{h}(\theta_B,R_B)}{\mathbf{h}^H(\theta_B,R_B)}]^2\}} \notag \\
&=\eta(1-\mathbb{E}_{k_n}\left[|\rho(q,p,k_n)|^2\right]),
\end{align}
where $\mathbf{P}(\theta_B,R_B)\!\triangleq\!\mathbf{I}_N-{\mathbf{h}(\theta_B,R_B)}{\mathbf{h}^H(\theta_B,R_B)}$. Noting that $\mathbf{z} \sim  \mathcal{CN}(0,\mathbf{I}_N)$,  and $\overset{b}{=}$ is obtained based on $\mathbb{E}[\mathbf{z}\mathbf{z}^H]=\mathbf{I}_N$.

Following \eqref{inequality}, after some algebraic manipulations we obtain the lower bound as given in \eqref{lower bound of ergodic capacity}, which completes the proof of this theorem.
\end{IEEEproof}

We note that the results provided in Theorem~\ref{theorem1} is valid for arbitrary values of $N$. Due to the distance concentration phenomenon \cite{francois2007}, we know that $|\rho(q, p, k_n)|^2$ approaches its mean $\mathbb{E}_{k_n}\left[|\rho(q,p,k_n)|^2\right]$ when $N \rightarrow \infty$, i.e., $|\mathbf{h}^H(\theta_E,R_E)\mathbf{h}(\theta_B,R_B)|^2$ in \eqref{gamma E} approaches its mean $\mathbb{E}_{k_n}\left[|\mathbf{h}^H(\theta_E,R_E)\mathbf{h}(\theta_B,R_B)|^2\right]$ when $N \rightarrow \infty$. As such, we can conclude that the lower bound approaches the ESC when $N \rightarrow \infty$. Therefore, we next determine the expression of the ESC when $N \rightarrow \infty$ in the following corollary.

\begin{corollary}\label{cor:lower bound of ESC}
As $N \rightarrow \infty$, the asymptotic ESC of the RFDA-DM-AN scheme is
\begin{multline}\label{asymptotic}
C_{\infty} =\\\log_2\left(\frac{\!-\!\alpha^2\mu_B^2 +\alpha\mu_B(\beta F_{\infty}+\mu_B\!-\!1)+\beta F_{\infty}+\mu_B}{\alpha\mu_B(F_{\infty}-\frac{1}{\eta}-1)+\beta F_{\infty}+\mu_B}\right),
\end{multline}
where
\begin{align}
F_{\infty}\triangleq \frac{N^2}{\eta(N^2-S^2_N(q)\Phi^2(j2\pi p))}.
\end{align}
\end{corollary}

\begin{IEEEproof}
As $N \rightarrow \infty$, we will have $\mathbb{V}_{k_n}[|\rho(q,p,k_n)|] \rightarrow 0$ due to the distance concentration phenomenon \cite{francois2007}. As such, following \eqref{mean of rho_square} we have
\begin{align}\label{mean_asymptotic}
\mathbb{E}_{k_n}\left[|\rho(q,p,k_n)|^2\right] &=\mathbb{E}_{k_n}^2\left[\rho(q,p,k_n)\right]\notag \\
&=\frac{1}{N^2}S^2_N(q)\Phi^2(j2\pi p).
\end{align}
Then, substituting \eqref{mean_asymptotic} into \eqref{inequality} and performing some algebraic manipulations, we can obtain the asymptotic ESC as given in \eqref{asymptotic}.
This completes the proof of Corollary~\ref{cor:lower bound of ESC}.
\end{IEEEproof}

Similar to \eqref{average capacity}, we can determine the average value of $C_{LB}$ over all possible locations of Eve as
\begin{align}
\overline{C}_{LB}=\int_{R_B\in \mathcal{R}}\int_{\theta_B\in\Theta} f(\theta_B,R_B) C_{LB} d\theta_B d R_B.
\end{align}
Then, the optimal value of $\alpha$ that maximizes $\overline{C}_{LB}$ can be obtained through
\begin{align}\label{optimal power allocation factor of capacity lower bound}
\alpha_{LB}^{\ast} = \underset{0\leq\alpha\leq1}{\mathrm{arg}\max}~~\overline{C}_{LB}.
\end{align}

In order to fully examine the benefits of the RFDA-DM-AN scheme, we derive the secrecy capacities of the DM with AN based on PA (referred to as the PA-DM-AN scheme) and the DM with AN based on LFDA (referred to as the LFDA-DM-AN scheme) as benchmarks in Appendix A.

\subsection{Continuous and Discrete Uniform Frequency Allocations}

We note that the lower bound derived in Theorem~\ref{theorem1} is valid for any MGF of $k_n$, i.e., for any random frequency allocation method. In this work, we consider the continuous uniform and discrete uniform frequency allocations, in which $k_n$ follows a continuous uniform distribution and a discrete uniform distribution, respectively.
The MGF of a continuous uniform random variable $t$ is given by
\begin{align}
\Phi(t)=\frac{e^{at}-e^{bt}}{t(a-b)},
\end{align}
where $t\in[a,b]$. Therefore, when $k_n$ is a continuous uniform random variable within $[-\frac{M}{2},\frac{M}{2}]$, its MGF is given by
\begin{align}\label{contineous}
\Phi(j2\pi p)&=\frac{e^{-\frac{M}{2}j2\pi p}-e^{\frac{M}{2}j2\pi p}}{j2\pi p(-\frac{M}{2}-\frac{M}{2})} \\ \nonumber
&=\frac{\sin(M\pi p)}{M\pi p},
\end{align}
where $M$ is determined by the total available frequency bandwith for the antenna array at Alice.

The MGF of a discrete uniform random variable $t$ is given by
\begin{align}
\Phi(t)=\frac{e^{at}-e^{(b+1)t}}{K(1-e^t)},
\end{align}
where $K$ is the number of all possible values of $t$ subject to $t \in [a,b]$. As such, when $k_n$ is within a discrete uniform set $\{-\frac{M-1}{2}, -\frac{M+1}{2}, \dots ,\frac{M-1}{2}\}$, its MGF is given by
\begin{align}\label{discrete}
\Phi(j2\pi p)&=\frac{e^{-\frac{M-1}{2}j2\pi p}-e^{(\frac{M-1}{2}+1)j2\pi p}}{M(1-e^{j2\pi p})}  \notag \\
&=\frac{\sin(M\pi p)}{M\sin(\pi p)}.
\end{align}

By substituting \eqref{contineous} and \eqref{discrete} into Theorem~\ref{theorem1}, we can obtain the lower bound on the secrecy capacity $C$ for the continuous uniform frequency allocation and discrete uniform frequency allocation, respectively. Accordingly, we can obtain the secrecy performance of these two frequency allocations, which will be examined in the following section.

\section{Numerical Results}

In this section, we numerically evaluate the secrecy performance of the RFDA-DM-AN (i.e., random frequency diverse array based directional modulation with artificial noise) scheme with the PA-DM-AN and LFDA-DM-AN schemes as benchmarks. Without other statements, our system settings used in this section are as follows. The carrier frequency $f_c$ is set to 1~GHz (i.e., $f_c = 1$GHz), the frequency increment is set to 3~MHz (i.e., $\Delta{f} = 3$MHz), the element spacing is half of the wavelength (i.e., $d=c/2f_c$), the location of Bob is set at ($45^{\circ}$, $120~\mathrm{m}$), and $\beta = 1$.

\begin{figure}
  \centering
  \includegraphics[width=3.2in, height=2.7in]{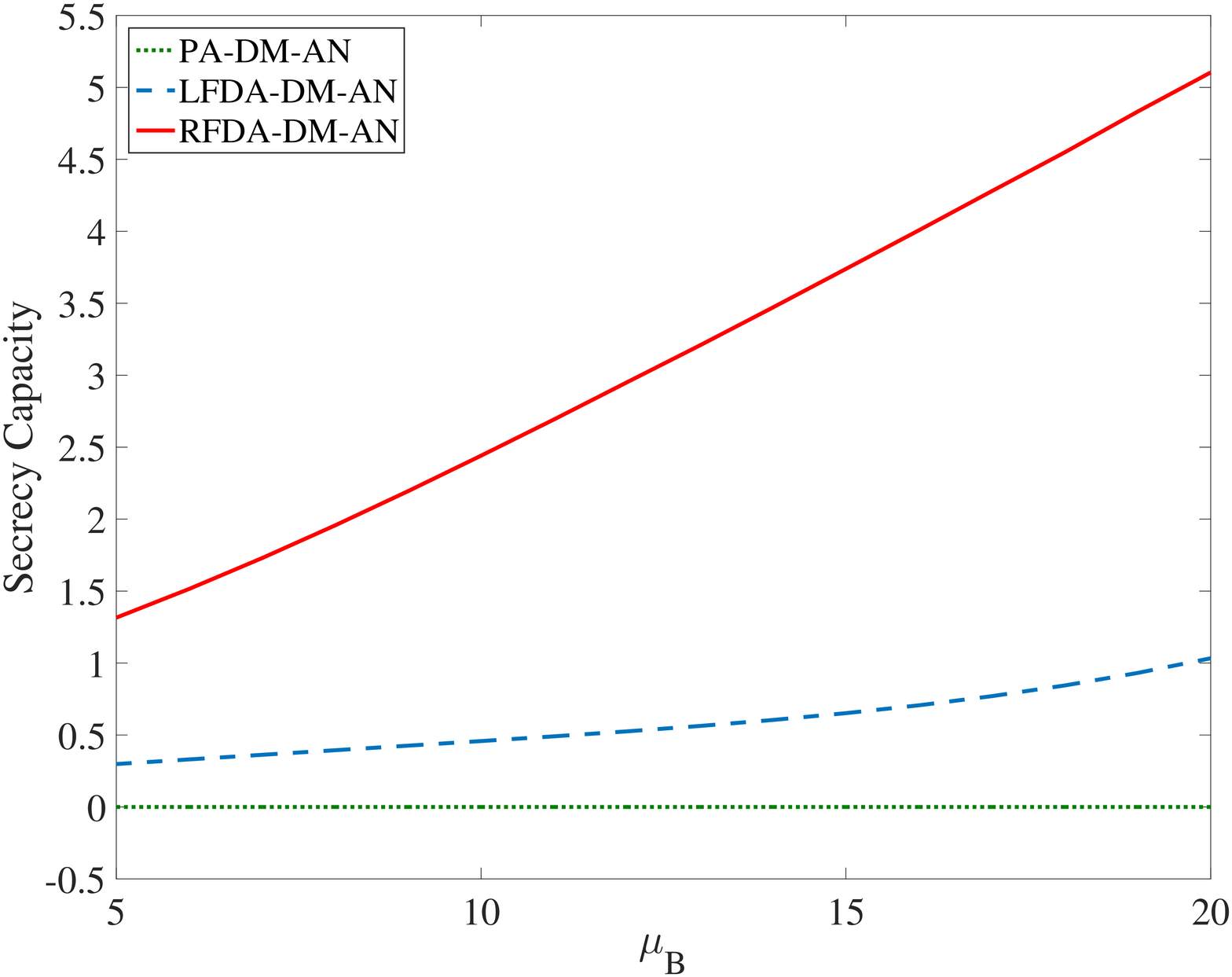}
  \caption{The ergodic secrecy capacity of the RFDA-DM-AN scheme and secrecy capacities of the PA-DM-AN and LFDA-DM-AN schemes versus $\mu_B$, where $N = 32$, Eve's location is ($45^{\circ}$, $239~\mathrm{m}$), and $\alpha = 0.5$.}\label{fig1}
\end{figure}

\begin{figure}
\begin{minipage}{0cm}
\includegraphics[width=3.6in, height=3in]{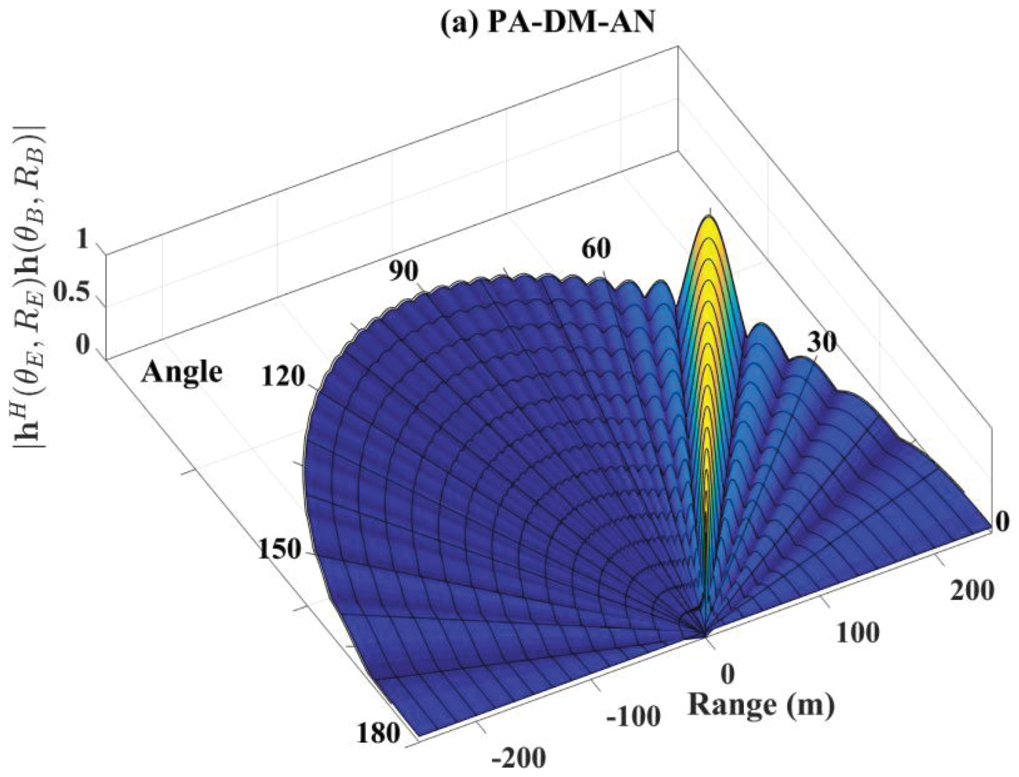}
\end{minipage}

\begin{minipage}{0cm}
\includegraphics[width=3.6in, height=3in]{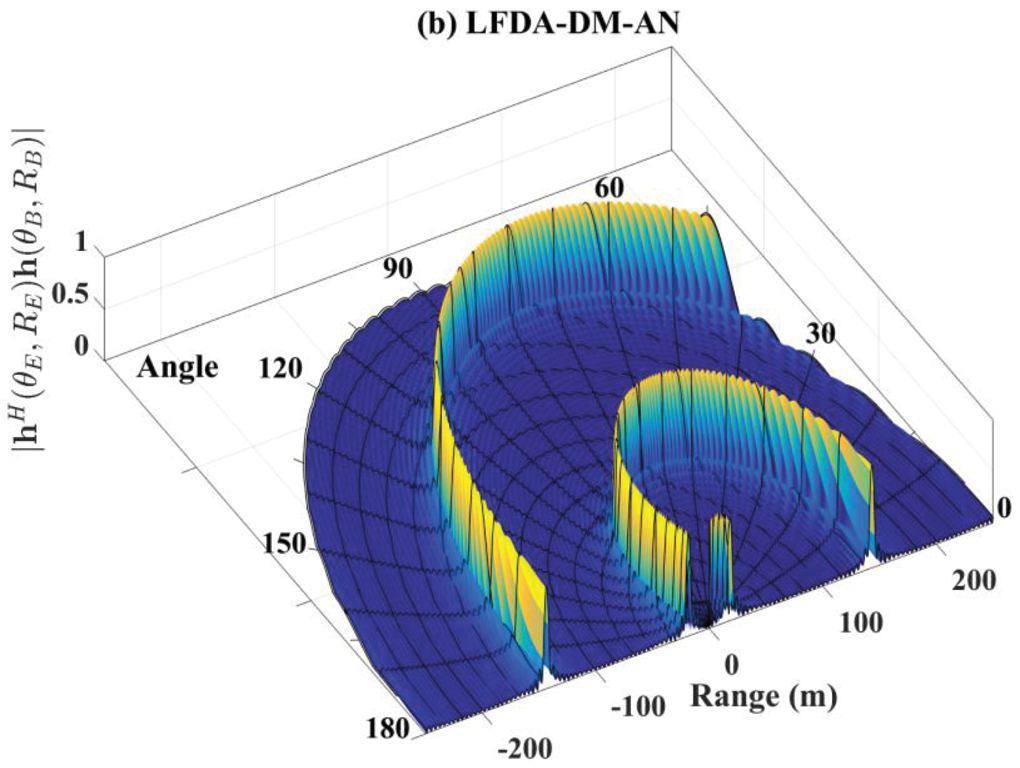}
\end{minipage}

\begin{minipage}{0cm}
\includegraphics[width=3.6in, height=3in]{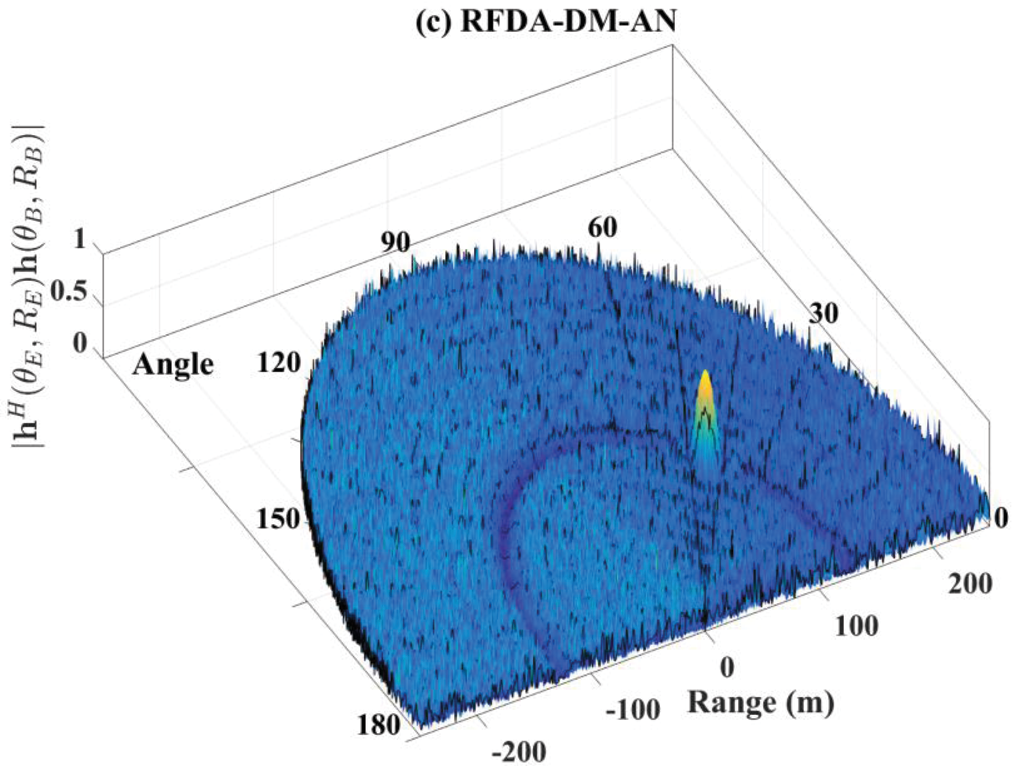}
\end{minipage}
\caption{$|\mathbf{h}^H(\theta_E,R_E)\mathbf{h}(\theta_B,R_B)|$ of the RFDA-DM-AN, PA-DM-AN, and LFDA-DM-AN schemes, where $N = 32$.}
\label{fig2}
\end{figure}

In Fig.~\ref{fig1}, we plot the ESC of the RFDA-DM-AN scheme and secrecy capacities of the PA-DM-AN and LFDA-DM-AN schemes versus $\mu_B$ for a specific location of Eve. We note that this Eve's location is only for the performance examination, which is unknown to Alice. As expected, we first observe that the secrecy capacity of the PA-DM-AN scheme is zero since Eve is in the same direction as Bob relative to Alice. In addition, we observe that the secrecy capacity of the LFDA-DM-AN scheme is much lower than the ESC of the RFDA-DM-AN scheme. This indicates that our proposed RFDA-DM-AN scheme can significantly outperform both the PA-DM-AN and LFDA-DM-AN schemes. We note that Eve can select the locations (not the same as Bob's location) that guarantee a zero secrecy capacity in the PA-DM-AN and LFDA-DM-AN schemes, since Eve may know Alice's location in practice. However, there are no such locations that Eve can select to ensure a zero ESC in the RFDA-DM-AN scheme. This is due to the fact that RFDA can decouple the correlation between the range and angle in DM, which is detailed in the following figure.

In Fig.~\ref{fig2}, we plot the absolute value of the correlation coefficient between $\mathbf{h}(\theta_E,R_E)$ and $\mathbf{h}(\theta_B,R_B)$ in the PA-DM-AN, LFDA-DM-AN, and RFDA-DM-AN schemes in order to further explain the observations found in Fig.~\ref{fig1}. As we can see from Fig.~\ref{fig2} (a), the maximum values (i.e., yellow areas) of this coefficient $|\mathbf{h}^H(\theta_E,R_E)\mathbf{h}(\theta_B,R_B)|$ appear in the direction of Bob, which means that if the eavesdropper exists along the desired direction, the secrecy capacity is zero (i.e., the received signals at Bob and Eve are identical). This explains why the secrecy capacity of the PA-DM-AN scheme is zero in Fig.~\ref{fig1}.
As shown in Fig.~\ref{fig2} (b), the maximum values of $|\mathbf{h}^H(\theta_E,R_E)\mathbf{h}(\theta_B,R_B)|$ appear periodically around Bob's location, which demonstrates that the range and angle are coupled in this scheme. The periodical peak values indicate that the LFDA-DM-AN scheme may not achieve positive secrecy capacity even when Eve is not at the same location. In Fig.~\ref{fig2} (c), we observe that the unique maximum value of $|\mathbf{h}^H(\theta_E,R_E)\mathbf{h}(\theta_B,R_B)|$ only occurs at the location of Bob, which means that a positive ESC can be achieved as long as Eve is not at the location of Bob. In practice, if Eve locates at the same location as Bob, Bob can inform Alice about this information in order to avoid Eve's attacks. As such, the aforementioned observations intuitively demonstrate the advantages of the RFDA-DM-AN scheme.

\begin{figure}
  \centering
  \includegraphics[width=3.2in, height=2.7in]{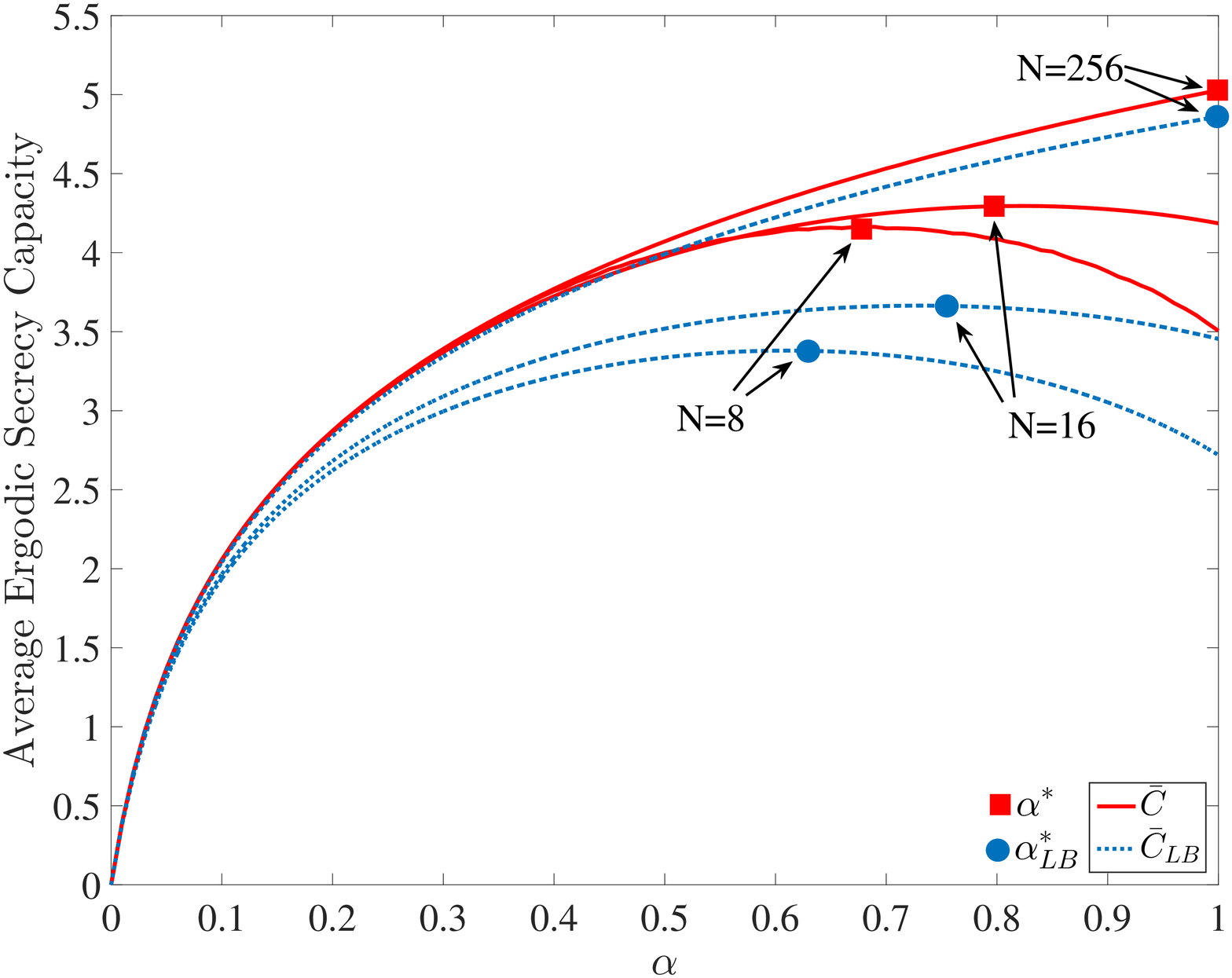}
  \caption{$\bar{C}$ and $\bar{C}_{LB}$ of the RFDA-DM-AN scheme versus $\alpha$, where $\mu_B = 15$dB.}\label{fig3}
\end{figure}

\begin{figure}
  \centering
  \includegraphics[width=3.2in, height=2.7in]{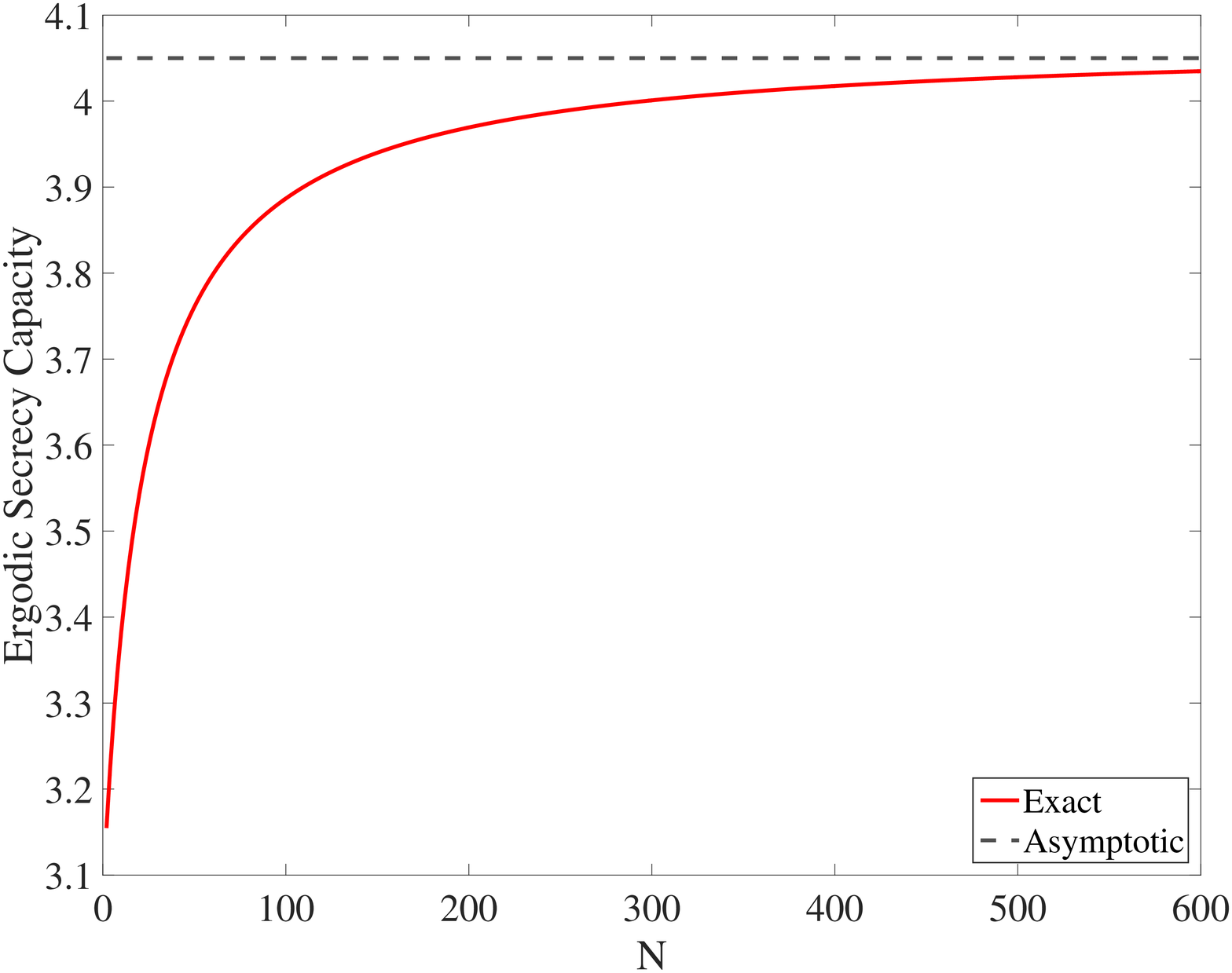}
  \caption{The exact and asymptotic ergodic secrecy capacity versus the different values of $N$, where $\mu_B = 15$dB.}\label{fig4}
\end{figure}

In Fig.~\ref{fig3}, we plot the average value of the ESC, i.e., $\overline{C}$, and the average value of the lower bound on the ESC, i.e., $\overline{C}_{LB}$, versus $\alpha$. For this figure, the potential location of Eve is uniformly distributed at $[0^\circ, {44}^\circ]\bigcup[{46}^\circ, {180}^\circ]$ in angle and $[0~\mathrm{m},119~\mathrm{m}]\bigcup[121~\mathrm{m},250~\mathrm{m}]$ in range. From Fig.~\ref{fig3}, we first observe that the gap between $\overline{C}$ and $\overline{C}_{LB}$ decreases as the number of antennas at Alice (i.e., N) increases. When $N$ is sufficiently large (e.g., $N=256$), we can see $\overline{C}_{LB}$ precisely matches $\overline{C}$, which can be explained by our Corollary~1. When $N$ is not very large (e.g., $N = 16$), we can see that the optimal value of $\alpha$ determined based on $\overline{C}_{LB}$ is still close to that determined based on $\overline{C}$. This demonstrates the validity of using $\overline{C}_{LB}$ as an approximation of $\overline{C}$ to determine the transmit power allocation between the useful signal and AN at Alice. Finally, in this figure we observe that the optimal value of $\alpha$ approaches one as $N$ increases. This demonstrates that Alice does not have to transmit AN when $N$ is sufficiently large, which is due to the fact that Alice can construct an ultra-narrow beam towards Bob when $N$ is large enough to avoid information leakage to Eve.
In Fig.~\ref{fig4}, we plot the exact and asymptotic ESCs versus different values of $N$. In this figure, we can observe that the exact ESC approaches the asymptotic one as $N$ increases, which confirms our Corollary~1.

\begin{figure}
\centering
\includegraphics[width=3.2in, height=2.7in]{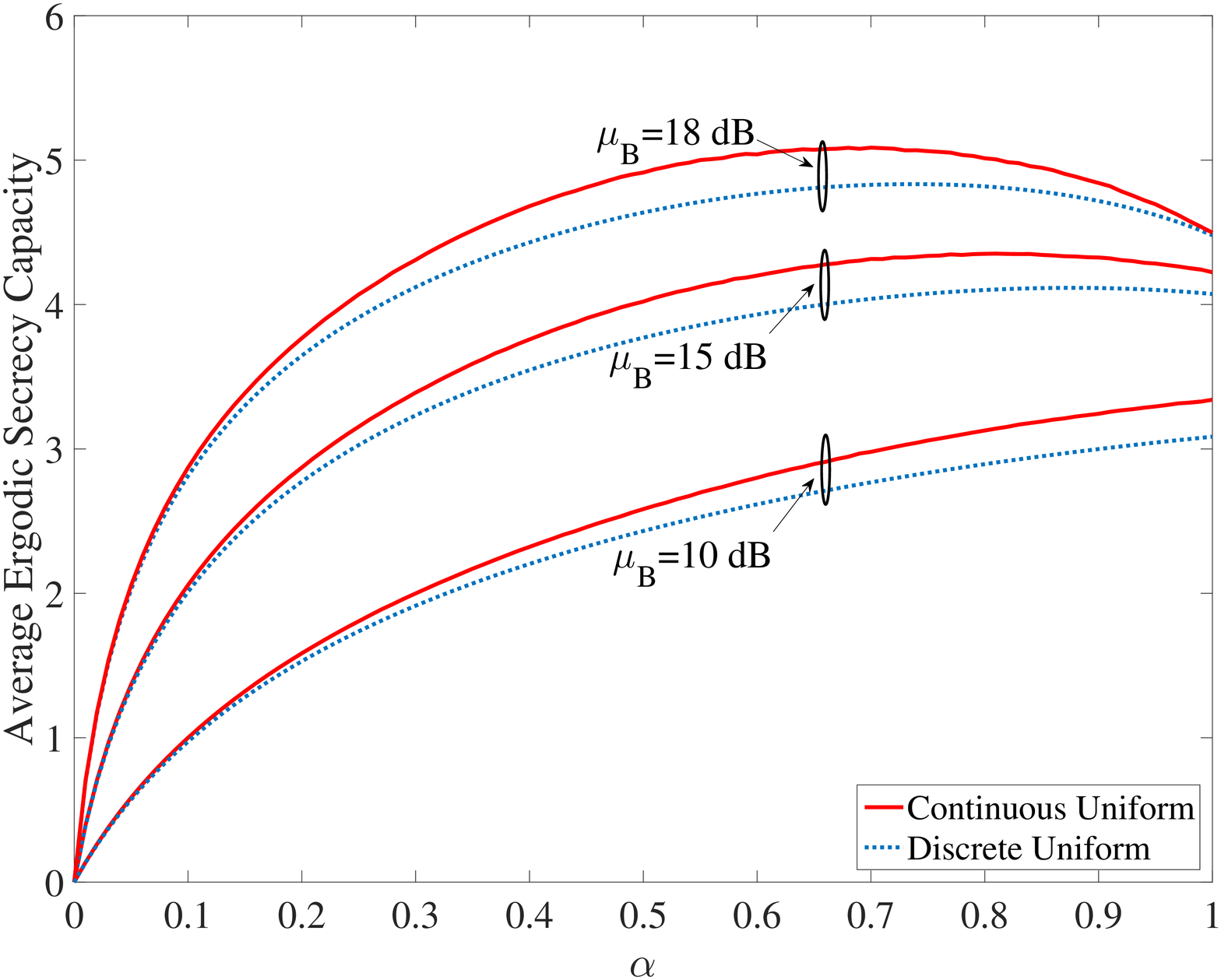}
\caption{Average ergodic secrecy capacity of the RFDA-DM-AN scheme with continuous and discrete uniform frequency allocations, where $N = 16$ and $M = 10$.}
\label{fig5}
\end{figure}

In Fig.~\ref{fig5}, we examine the secrecy performance of the RFDA-DM-AN scheme with continuous and discrete uniform frequency allocations. As we can see from Fig.~\ref{fig5}, the continuous uniform frequency allocation outperforms the discrete one in terms of average ESC. In this figure, we also observe that the average ESC increases as $\mu_B$ increases, which indicates that Alice can enhance physical layer security through increasing her transmit power. Finally, we observe that the optimal value of $\alpha$ that maximizes the average ESC increases as $\mu_B$ decreases. This indicates that Alice allocates a larger fraction of her transmit power to the useful signal as her transmit power decreases, and she allocates all her transmit power to the useful signal (i.e., $\alpha = 1$) when her transmit power is sufficiently low as shown in Fig.~\ref{fig5}.

\section{Conclusion}

In this work, we proposed the RFDA-DM-AN scheme to enhance physical layer security of wireless communications. We derived a lower bound on the ESC of the proposed scheme, based on which the transmit power allocation between the useful signal and AN was efficiently determined. We also derived an asymptotic ESC when $N$ approaches infinity, which precisely matches our derived lower bound when $N$ is sufficiently large.
Our investigation demonstrates that the proposed RFDA-DM-AN scheme can significantly outperform the PA-DM-AN and LFDA-DM-AN schemes in terms of average ESC.

\appendices

\section{Secrecy Capacities of the PA-DM-AN and LFDA-DM-AN Schemes}\label{app:LFDA and PA}

For PA, the phase shift of the $n$-th element relative to the reference element is given by \cite{Ding2}
\begin{align}\label{phase shift of PA}
\Psi_n(\theta)=\frac{2\pi b_n}{c}(-{f_c d\cos\theta}).
\end{align}
As a result, the cross correlation coefficient between $\mathbf{h}(\theta_E,R_E)$ and $\mathbf{h}(\theta_B,R_B)$ for PA is given by
\begin{align}
\mathbf{h}^H(\theta_E,R_E)\mathbf{h}(\theta_B,R_B)&=\frac{1}{N}\sum^{N-1}_{n=0}\Psi_n(\theta) \notag \\
&=\frac{1}{N}\frac{\sin(N\pi q)}{\sin(\pi q)}.
\end{align}

Since the frequency allocation to each antenna element in PA is fixed, the instantaneous secrecy capacity of the PA-DM-AN can be written as
\begin{align}
\label{cappacity of PA}
C&=\log_2(1+\alpha \mu_B) \notag \\
&~~-\log_2\left(1+\frac{\alpha\mu_B(\frac{\sin^2(N \pi p)}{N^2\sin^2(\pi p)})}{(1-\alpha)\mu_B \eta [1-\frac{\sin^2(N \pi p)}{N^2\sin^2(\pi p)}]+\beta}\right).
\end{align}

For LFDA, the phase shift of the $n$-th element relative to the reference element is given by \cite{Sammartino}
\begin{align}\label{phase shift of LFDA}
\Psi_n(\theta,R)=\frac{2\pi b_n}{c}(-{f_c d\cos\theta}+{\Delta{f}R}).
\end{align}

Following \eqref{phase shift of LFDA}, the cross correlation coefficient between $\mathbf{h}(\theta_E,R_E)$ and $\mathbf{h}(\theta_B,R_B)$ of LFDA is given by
\begin{align}
\label{cross correlation coefficient of these two steering vector LFDA}
\mathbf{h}^H(\theta_E,R_E)\mathbf{h}(\theta_B,R_B)&=\frac{1}{N}\sum^{N-1}_{n=0}\Psi_n(\theta,R) \notag \\
&=\frac{1}{N}\frac{\sin(N\pi(q-p))}{\sin(\pi(q-p))}.
\end{align}

Substituting \eqref{cross correlation coefficient of these two steering vector LFDA} into \eqref{gamma E}, we can obtain the SINR at Eve. Accordingly, the instantaneous secrecy capacity of the LFDA-DM-AN is given by
\begin{align}
\label{cappacity of LFDA}
C&=\log_2(1+\alpha \mu_B) \notag \\
&~~~-\log_2\left(1+\frac{\alpha\mu_B(\frac{\sin^2[N \pi(p-q)]}{N^2\sin^2[\pi (p-q)]})}{(1-\alpha)\mu_B \eta [1-\frac{\sin^2[N \pi(p-q)]}{N^2\sin^2[\pi (p-q)]}]+\beta}\right).
\end{align}

\bibliographystyle{IEEEtran}
\bibliography{IEEEfull,DM}

\begin{thebibliography}{10}
\providecommand{\url}[1]{#1}
\csname url@samestyle\endcsname
\providecommand{\newblock}{\relax}
\providecommand{\bibinfo}[2]{#2}
\providecommand{\BIBentrySTDinterwordspacing}{\spaceskip=0pt\relax}
\providecommand{\BIBentryALTinterwordstretchfactor}{4}
\providecommand{\BIBentryALTinterwordspacing}{\spaceskip=\fontdimen2\font plus
\BIBentryALTinterwordstretchfactor\fontdimen3\font minus
  \fontdimen4\font\relax}
\providecommand{\BIBforeignlanguage}[2]{{%
\expandafter\ifx\csname l@#1\endcsname\relax
\typeout{** WARNING: IEEEtran.bst: No hyphenation pattern has been}%
\typeout{** loaded for the language `#1'. Using the pattern for}%
\typeout{** the default language instead.}%
\else
\language=\csname l@#1\endcsname
\fi
#2}}
\providecommand{\BIBdecl}{\relax}
\BIBdecl

\bibitem{yangnan2016}
N.~Yang, L.~Wang, G.~Geraci, M.~Elkashlan, J.~Yuan, and M.~D. Renzo,
  ``Safeguarding 5g wireless communication networks using physical layer
  security,'' \emph{IEEE Commun. Mag.}, vol.~53, no.~4, pp. 20--27, Apr. 2015.

\bibitem{Trappe2015}
W.~Trappe, ``The challenges facing physical layer security,'' \emph{IEEE
  Commun. Mag.}, vol.~53, no.~6, pp. 16--20, Jun. 2015.

\bibitem{hebiao2016}
B.~He, X.~Zhou, and T.~D. Abhayapala, ``Wireless physical layer security with
  imperfect channel state information: A survey,'' \emph{ZTE Commun.}, vol.~11,
  no.~3, pp. 11--19, Sep 2013.

\bibitem{yan2014transmit}
S.~Yan, N.~Yang, R.~Malaney, and J.~Yuan, ``Transmit antenna selection with
  alamouti coding and power allocation in mimo wiretap channels,'' \emph{IEEE
  Trans. Wireless Commun.}, vol.~13, no.~3, pp. 1656--1667, Mar. 2014.

\bibitem{long2}
Y.~L. Zou, J.~Zhu, X.~Wang, and V.~Leung, ``Improving physical-layer security
  in wireless communications through diversity techniques,'' \emph{IEEE Net.},
  vol.~29, no.~1, pp. 42--48, Jan. 2015.

\bibitem{yan2016location}
S.~Yan and R.~Malaney, ``Location-based beamforming for enhancing secrecy in
  rician wiretap channels,'' \emph{IEEE Trans. Wireless Commun.}, vol.~15,
  no.~4, pp. 2780--2791, Apr. 2016.

\bibitem{chenxiaoming}
X.~Chen, D.~W.~K. Ng, and H.~Chen, ``Secrecy wireless information and power
  transfer: challenges and opportunities,'' \emph{IEEE Wireless Commun.},
  vol.~23, no.~2, pp. 54--61, Apr. 2016.

\bibitem{zhaonan1}
N.~Zhao, F.~R. Yu, M.~Li, and V.~C.~M. Leung, ``Anti-eavesdropping schemes for
  interference alignment ({I}{A})-based wireless networks,'' \emph{IEEE Trans.
  Wireless Commun.}, vol.~15, no.~8, pp. 5719--5732, Aug. 2016.

\bibitem{Babakhani1}
A.~Babakhani, D.~Rutledge, and A.~Hajimiri, ``Transmitter architectures based
  on nearfield direct antenna modulation,'' \emph{IEEE J. Solid-State
  Circuits}, vol.~43, no.~12, pp. 2674--2692, Dec. 2008.

\bibitem{Babakhani2}
------, ``Near-field direct antenna modulation,'' \emph{IEEE Microw. Mag.},
  vol.~10, no.~1, pp. 36--46, Feb. 2009.

\bibitem{Daly1}
M.~P. Daly and J.~T. Bernhard, ``Directional modulation technique for phased
  arrays,'' \emph{IEEE Trans. Antennas Propag.}, vol.~57, no.~9, pp.
  2633--2640, Sep. 2009.

\bibitem{Daly2}
M.~P. Daly, E.~L. Daly, and J.~T. Bernhard, ``Demonstration of directional
  modulation using a phased array,'' \emph{IEEE Trans. Antennas Propag.},
  vol.~58, no.~5, pp. 1545--1550, May 2010.

\bibitem{Ding2}
Y.~Ding and V.~Fusco, ``A vector approach for the analysis and synthesis of
  directional modulation transmitters,'' \emph{IEEE Trans. Antennas Propag.},
  vol.~62, no.~1, pp. 361--370, Jan. 2014.

\bibitem{Jinsong}
J.~Hu, F.~Shu, and J.~Li, ``Robust synthesis method for secure directional
  modulation with imperfect direction angle,'' \emph{IEEE Commun. Lett.},
  vol.~20, no.~6, pp. 1084--1087, Jun. 2016.

\bibitem{Shufeng}
F.~Shu, X.~Wu, J.~Li, R.~Chen, and B.~Vucetic, ``Robust synthesis scheme for
  secure multi-beam directional modulation in broadcasting systems,''
  \emph{IEEE Access}, vol.~4, pp. 6614--6623, Oct. 2016.

\bibitem{Antonik:PHDthesis}
P.~Antonik, ``An investigation of a frequency diverse array,'' Ph.D.
  dissertation, University College London, London, UK, 2009.

\bibitem{Sammartino}
P.~Sammartino, C.~Baker, and H.~Griffiths, ``Frequency diverse mimo techniques
  for radar,'' \emph{IEEE Trans. Aerosp. Electron. Syst.}, vol.~49, no.~1, pp.
  201--222, Jan. 2013.

\bibitem{wangwenqin2015}
W.~Q. Wang, ``Frequency diverse array antenna: New opportunities,'' \emph{IEEE
  Antennas Propag. Mag.}, vol.~57, no.~2, pp. 145--152, Apr. 2015.

\bibitem{Liu1}
Y.~Liu, ``Range azimuth indication using a random frequency diverse array,'' in
  \emph{Proc. IEEE Int. Conf. on Acoustics, Speech and Signal Process. (ICASSP
  2016)}, Mar. 2016, pp. 3111--3115.

\bibitem{Liu2}
Y.~Liu, H.~Rui, L.~Wang, and A.~Nehorai, ``The random frequency diverse array:
  a new antenna structure for uncoupled direction-range indiction in active
  sensing,'' \emph{IEEE Jour. on Sel. Topics in Signal Process.}, to apear,
  2016.

\bibitem{Goel}
S.~Goel and R.~Negi, ``Guaranteeing secrecy using artificial noise,''
  \emph{IEEE Trans. Wireless Commun.}, vol.~7, no.~6, pp. 2180--2189, Jun.
  2008.

\bibitem{yangnan2015}
N.~Yang, S.~Yan, J.~Yuan, R.~Malaney, R.~Subramanian, and I.~Land, ``Artificial
  noise: Transmission optimization in multi-input single-output wiretap
  channels,'' \emph{IEEE Trans. Commun.}, vol.~63, no.~5, pp. 1771--1783, May
  2015.

\bibitem{Shihao}
S.~Yan, N.~yang, G.~Geraci, R.~Malaney, and J.~Yuan, ``Optimization of code
  rates in sisome wiretap channels,'' \emph{IEEE Trans. Wireless Commun.},
  vol.~14, no.~11, pp. 6377--6388, Nov. 2015.

\bibitem{francois2007}
D.~Francois, V.~Wertz, and M.~Verleysen, ``The concentration of fractional
  distances,'' \emph{IEEE Trans. on Knowl. Data. Eng.}, vol.~19, no.~7, pp.
  873--886, Jul. 2007.

\end{thebibliography}

\end{document}